\documentclass{osa-article}
\journal{oe}

\articletype{Research Article}

\begin{document}

\title{K-series approximation of vectorial optical fields for designing diffractive optical elements with subwavelength feature sizes}

\author{I-Lin Ho\authormark{*}}
\address{Institute of Nuclear Energy Research, Atomic Energy Council, Executive Yuan, Taoyuan City 32546, Taiwan (ROC)}
\email{\authormark{*}suntaho@iner.gov.tw}


\begin{abstract}
Diffractive optical elements (DOEs) are widely applied as compact solutions for desired light manipulations via wavefront shaping. Recent advanced chip applications further require their feature sizes to move down to the subwavelength, which inevitably brings forth vectorial effects of optical fields and makes the typical scalar-based theory invalid. However, simulating and optimizing their vectorial fields, which are associated with billions of adjustable parameters in the optical element, are difficult to do, because of the issues of numerical stability and the highly-demanding computational cost. To address this problem, this research proposes an applicable algorithm by means of a wave-vector (k) series approximation of vectorial optical fields. On the basis of the semi-analytical rigorous coupled wave analysis (RCWA), an adequate selection scheme on k-series enables computationally efficient yet still predictive calculations for DOEs. The performance estimations for exemplary designs by the finite difference time domain (FDTD) method show that the predicted intensity profiles by the proposed algorithm agree with the target by just a fractional error. Together with optimizing the geometrical degrees of freedom (e.g., DOE depth h) as compensation for errors from the truncation of k-series, the algorithm demonstrates its outperformance by one or two orders of magnitude in accuracy versus the scalar-based model, and demands only a reasonable computational resource.
\end{abstract}



\section{Introduction}

Diffractive optical elements (DOEs) are devices composed of transparent materials with complex topography to realize specified functions (e.g. optical computing and cell trapping \cite{oapp1,oapp2}) that are not feasible with standard refractive optics. With the rapid development of additive manufacturing techniques \cite{3dp1,3dp2,3dp3} in recent years, especially two-photon polymerization lithography (TPL) \cite{tpl1,tpl2,tpl3}, it has become more feasible to fabricate DOEs with feature sizes down to ten nanometers. So far this technology is widely used for optical applications in the visible band, such as holograms for virtual reality \cite{app1}, colorful 3D prints \cite{app2}, and optical anticounterfeiting devices \cite{app3}. Moreover, TPL has been proven to bear potential at producing DOEs for advanced chip-scale applications like all-optical diffractive deep neural networks \cite{app4,app5,app6}. However, for the mentioned feature size, the assumptions of typical scalar-based design methods are violated \cite{scalar1,scalar2} and the vectorial effects of optical fields become more pronounced \cite{vec1}. A general, fully rigorous design tool is therefore needed to enable an accurate synthesis of such elements. So far this is quite a challenging work for researchers.

A small strand of the literature has reported on designing DOEs with subwavelength feature sizes. For DOEs in the 1D domain, D.W. Prather et al. proposed an optimization-based algorithm using the boundary element method (BEM) \cite{bem1}. Its computational costs, however, prevent the design of realistic DOEs in reasonable time frames. J. Jiang et al. developed the micro-genetic algorithm FDTD method \cite{vec2}, but it remains a challenging design tool due to its time-consuming searching routines. M.E. Testorf et al. demonstrated the Gerchberg-Saxton-algorithm FDTD method \cite{vec3}, which is a highly efficient optimization algorithm, but is related to the paraxial design algorithm. Feng Di et al. employed an iterative optimization algorithm \cite{vec4,vec5} based on the rigorous electromagnetic theory for a 1D beam splitter, but it is still subject to time-consuming and local-searching-only issues. In the 2D domain, H. Hao et al. presented a hybrid method that combines the rigorous coupled wave analysis (RCWA) \cite{rcwa1,rcwa2} and the genetic algorithm \cite{vec6} to design 2D wide-angle beam splitters. This method, however, precludes designs from DOE's applications other than splitters. S. J. Byrnes et al. demonstrated RCWA-based simulations to realize novel meta-lenses \cite{metalens1}. The technique takes advantage of the peripherally-radial periodicity and the circular symmetry of a lens to efficiently design the metasurfaces, but could treat other symmetries in an unfavorable way; e.g., C2 symmetry in binary DOEs or asymmetry in ternary DOEs. S. Wei et al. utilized optical properties of the unit cell (an elliptical silicon pillar in a $\mathrm{0.5\times0.5\times0.6 \ \mu m^{3}}$ box) to design the dielectric metasurface of the polarimeter system \cite{polarimeter1}, featuring high (above $\mathrm{60\%}$) detection efficiency for polarized light. This semi-analytical approach, however, could inherently suppress further improvement of the design due to strong phase gradients, interactions between neighboring pillars, and oblique angles \cite{metalens1}. N. Li et al. explored a large-area ($\mathrm{2500\mu m \times 2500 \mu m}$) metasurface beam deflector \cite{defl1}, showing the pixel level ($\mathrm{120\mu m \times 120 \mu m}$) controlling light. As discussed by the authors, the finite efficiency of the device needs to be improved by optimizing the design of the nanopillars, taking the interactions among nearby pillars with various diameters into consideration. Thus, it is an urgent requirement to develop an efficient algorithm for designs of DOEs that are finite in extent (mm- to cm-scale), have subwavelength features, and are aperiodic for different functions.

On the back of the scalar-based simulated annealing algorithm (s-SAA) \cite{sa1}, this research proposes a vector-based simulated annealing algorithm (v-SAA) for designing DOEs with subwavelength feature sizes. By considering the vectorial effects of optical fields dominatingly taking place inside the elements, this work mainly re-models the wave front function $\mathrm{U}$ in s-SAA with the rigorous electromagnetic theory. Specifically, on the basis of RCWA, a k-series approximation method is introduced to quickly calculate the pixel-field profile $\mathrm{\mathbf{P}}$ (in contrast to $\mathrm{U}$ in s-SAA) during the v-SAA process, where the order of k-series correlates with optional parameters for selecting different degrees of accuracy and computational costs. The v-SAA hence can achieve computationally efficient yet still predictive optical designs for DOEs.

The rest of the paper runs as follows. Four subsections of section 2 are arranged based on the following considerations: identify problems in the process, find solutions to problems, integrate the solutions into the process, and improve the new assembly (with new solution component) process. Specifically, section 2 describes the detailed design technique and points out the limits of the scalar-based theory on subwavelength-scale designs. To break the barriers of scalar-based inaccuracy and the vector-based high cost, a k-series approximation method (KSAM) is introduced to efficiently compute the vectorial optical fields of given DOE morphologies $\mathrm{\Upsilon}$. On the basis of KSAM, an iterative algorithm v-SAA is then realized for efficient designs of DOEs with subwavelength feature sizes. The complementary optimization process for errors from the truncation of k-series is finally mentioned for high-quality products. Section 3 presents the design results and analysis to demonstrate the effectiveness of the proposed method. Section 4 concludes this work.

\section{Design principle}
\subsection{Limits of the scalar-based theory}

\begin{figure}[htbp]
\centering
\includegraphics[width=0.95\linewidth]{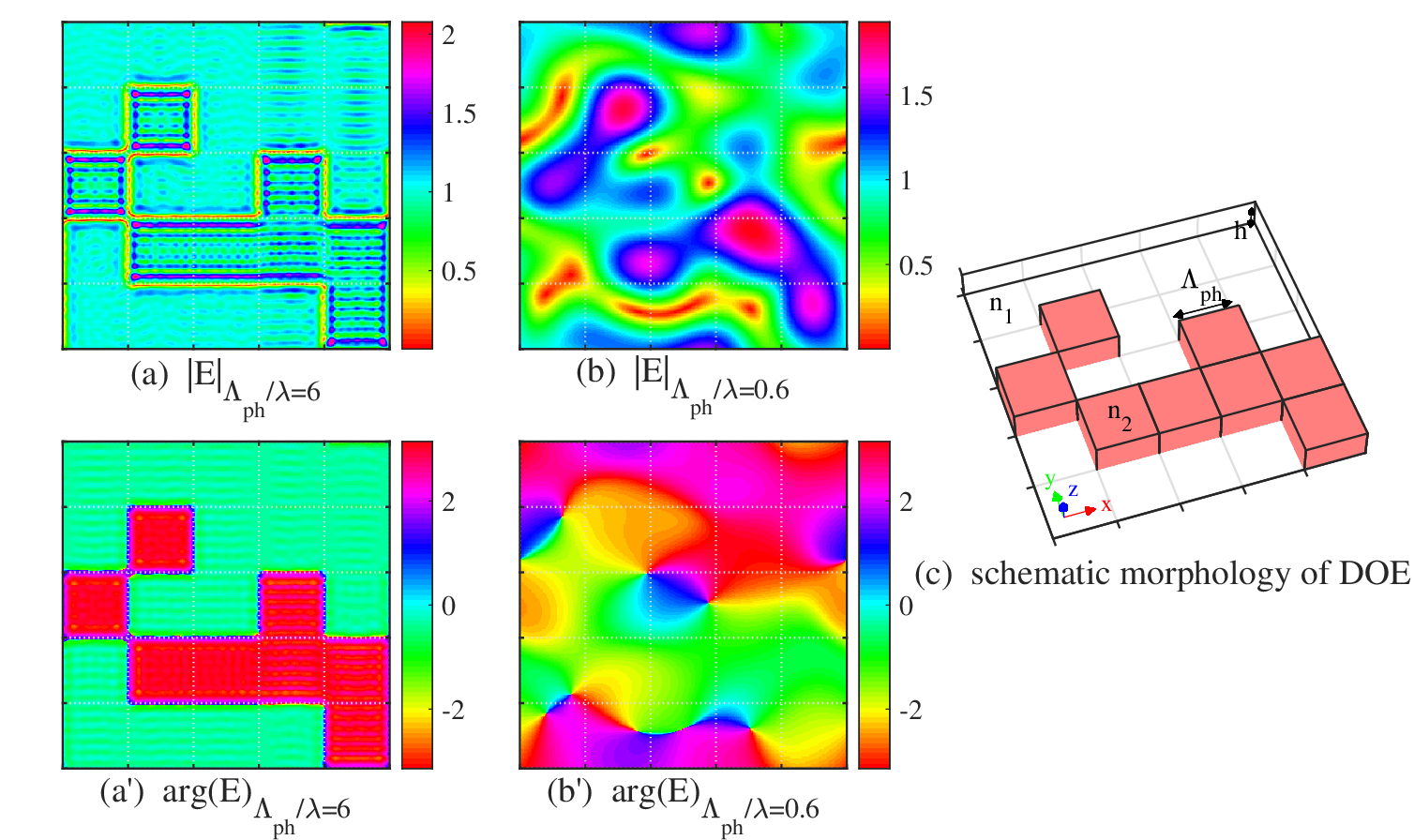}
\caption{(a)-(a') Amplitude and phase of fields for DOE with $\Lambda_\mathrm{ph}/\lambda$=6, (b)-(b') Amplitude and phase of fields for DOE with $\Lambda_\mathrm{ph}/\lambda$=0.6, (c) Schematic morphology of pixels of DOE.}
\label{fig1x}
\end{figure}

To interpret the deviation from the prediction by the scalar-based theory in systems with subwavelength feature size, in this subsection we apply the rigorous electromagnetic theory, FDTD, for accurate computations of vectorial optical fields. The vectorial effects are then demonstrated by observing an exemplary DOE at different feature sizes $\mathrm{\Lambda_{ph}}$. The DOE is composed of $5\times5$ square pixels with feature size $\mathrm{\Lambda_{ph}}$ and depth h, as indicated in Fig. \ref{fig1x}(c). Here two kinds of media, the substrate medium with refractive index $n_{2}=1.5$ and air with $n_{1}=1.0$, occupy the pink and transparent pixels, respectively. Figure \ref{fig1x}(a) illustrates the transmission fields on the exiting surface of DOE (at z=h+$\mathrm{\delta_{h}}$=$\mathrm{\lambda+\delta_{h}}$), providing a plane-wave incidence with wavelength $\lambda$ at $\mathrm{z=0}$ and structure parameter $\mathrm{\Lambda_{ph}/\lambda=6}$. An appropriate separation $\mathrm{\delta_{h}}$ from the surface $\mathrm{z=h}$ is adopted to diminish the influence of evanescent waves. In this classical regime of refractive optics- i.e., $\mathrm{\Lambda_{ph}/\lambda\gg1}$ - numerical results show that the intensity profile of optical fields is only dependent on the morphology (the pink or transparent pixel) of an individual pixel and has almost constant amplitude value $\mathrm{|E|=1}$. In Fig. \ref{fig1x}(a')  the corresponding phase profile $\mathrm{arg(E)}$ of the fields also shows significant dependence on the local pixel morphology, where the value of the phase difference between two kinds of pixels is $\mathrm{\pi}$, confirming the prediction by the scalar-based theory (see the texts below) \cite{sa1}. Compared to Figs. \ref{fig1x}(a) and \ref{fig1x}(a'), Figs. \ref{fig1x}(b) and \ref{fig1x}(b') clearly show that both the intensity and the phase of the transmission fields exhibit across-pixel continuity and display the sub-pixel profile for the DOE with $\mathrm{\Lambda_{ph}/\lambda=0.6}$. The assumption of uniformity and local-pixel-dependence of optical fields made by the scalar-based theory, i.e., $\mathrm{|E|=1}$ and $\mathrm{arg(E)=2\pi n_{1}h/\lambda}$ for the transparent pixel, and $\mathrm{|E|=1}$ and $\mathrm{arg(E)=2\pi n_{2}h/\lambda}$ for the pink pixel, is hence considerably invalid. Such results may also imply that these single-value representation of optical fields per pixel by the scalar-based theory can be problematic in the regime $\mathrm{\Lambda_{ph}/\lambda\leq1}$.

\begin{figure}[htbp]
\centering
\includegraphics[width=0.95\linewidth]{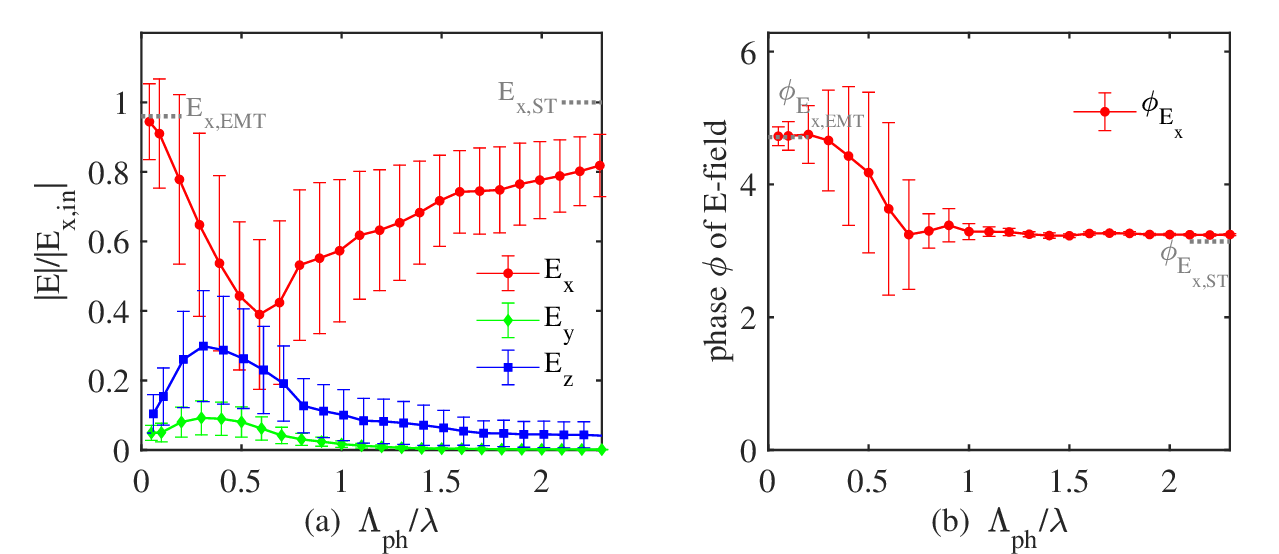}
\caption{(a) Normalized amplitude of different field components and (b) phase of the field's x-component at the central pixel as a function of pixel size by ensemble statistics. The FDTD method is adopted here for valid conclusions.}
\label{fig2x}
\end{figure}

To explore the generality of vectorial effects for systems studied, we statistically analyze fields of a specified (central) pixel inside DOE as a function of pixel size, by means of an ensemble of 20000 specimens per $\mathrm{\Lambda_{ph}}$ value. Similar to the schematic DOE in Fig. \ref{fig1x}(c), each specimen has its plane dimension $3.5\mathrm{\lambda\times3.5\lambda}$ and is assigned with stochastic pixel morphology except for the specified (central) one. Here, the FDTD method is adopted for giving convincing conclusions. Numerical results for amplitudes of different field components on the exiting surface (at z=$\mathrm{\lambda+\delta_{h}}$) are plotted in Fig. \ref{fig2x}(a), provided that the incident field has x polarization. The curve of phase of the field's x component is shown in Fig. \ref{fig2x}(b), where the curves for the other two components are not displayed, owing to their simple uniformly random distribution over the 0-2$\pi$ range. In principle, Fig. \ref{fig2x} portrays the properties of optical fields from the regime $\mathrm{\Lambda_{ph}/\lambda\gg1}$ governed by the scalar-based theory (ST), across the regime $0\ll\mathrm{\Lambda_{ph}/\lambda\leq1}$ belonging to vectorial optics, and to the regime $\mathrm{\Lambda_{ph}/\lambda\ll1}$ conforming to the effective medium theory (EMT)\cite{scalar2}. As expected, numerical results show that the amplitude and phase of fields exhibit asymptotic convergence toward constants $\mathrm{E_{x,EMT}}$ and $\phi_\mathrm{E_{x,EMT}}$ by EMT at $\mathrm{\Lambda_{ph}/\lambda\ll1}$, as well as toward constants $\mathrm{E_{x,ST}}$ and $\mathrm{\phi_{E_{x,ST}}}$ by ST at $\mathrm{\Lambda_{ph}/\lambda\gg1}$, respectively. Considerable depolarization and dephasing of fields, however, occur in the in-between regime and hence demand vector-based theories for different requirements.

\subsection{K-series approximation method to compute pixel fields $\mathrm{\mathbf{P}(x,y)}$ by the vectorial model}

To include vectorial effects into designs of finite-size (mm- to cm-scale) devices (with structure parameter $\mathrm{\Lambda _{ph}/\lambda \leq 1}$) for practical applications, the two most popular rigorous electromagnetic theories, FDTD and RCWA, have been adopted for various attempts as discussed in Sec. 1. Unfortunately, the FDTD method signifies highly-challenging issues, since its computation time scales are the fourth order of the simulation (spatial) domain size N (e.g., $\mathrm{(10N)^{4}\approx 10^{(1+5)\times 4}}$ at cm-scale), and the memory requirements are the third order \cite{fdtd0}. The RCWA method has the same issues due to its computation time, as well as memory requirements, being fourth order of the simulation (momentum space) domain size K \cite{rcwa3} (e.g., $\mathrm{K^{4}\approx 10^{5\times 4}}$ at cm-scale). To efficiently design optics of DOEs, on the basis of the semi-analytical RCWA framework, this subsection presents a k-series approximation method associated with optional parameters, the rank of neighboring interferences r, and the sampling rate s, for selecting desired degrees of accuracy at the cost of computational resources. In principle, the method scales the computation time and memory requirement down to the level of $\mathrm{N^{2}(2r+1)^{2}s^{4}}$; e.g., about $\mathrm{10^{11}}$ at condition r=s=1 at cm-scale, in comparison with the level of $\mathrm{N^2}\simeq 10^{10}$ by the scalar-based method \cite{sa1}. Here, the translation symmetry has been considered to estimate the computational cost for KSAM. We note that, implemented with lookup table techniques, the proposed method can further scale the computation time down to the level of $\mathrm{N^{2}}$; whereas too large values of r and s can burden this method with high computational costs beyond FDTD and RCWA. The underlying criteria are described in detail in the following paragraphs.

For a binary DOE (one stacking layer), its permittivity and the incident field can be represented in a Fourier series of spatial harmonics \cite{rcwa1} according to
\begin{eqnarray}
\mathrm{\varepsilon (x,y)} &=&\mathrm{\sum_{u,v=-\infty }^{\infty}\varepsilon _{uv}exp\left[ i\frac{2\pi ux}{\Lambda _{x}}+i\frac{2\pi vy}{\Lambda _{y}}\right] ,}  \label{ep1} \\
\mathrm{\varepsilon ^{-1}(x,y)} &=&\mathrm{\sum_{u,v=-\infty }^{\infty }\overline{\varepsilon }_{uv}exp\left[ i\frac{2\pi ux}{\Lambda _{x}}+i\frac{2\pi vy}{\Lambda _{y}}\right] ,}  \label{ep2} \\
\mathrm{\mathbf{E}} &=&\mathrm{\sum_{u,v=-\infty }^{\infty }\mathbf{e}_{uv}exp\left[ -ik_{xu}x-ik_{yv}y\right] ,}  \label{ep3} \\
\mathrm{\mathbf{H}} &=&\mathrm{-i\sqrt{\frac{\varepsilon _{0}}{\mu _{0}}}\sum_{u,v=-\infty }^{\infty }\mathbf{h}_{uv}exp\left[ -ik_{xu}x-ik_{yv}y\right] ,}  \label{ep4}
\end{eqnarray}%
in which the wave vector components $\mathrm{k_{xu}}$ and $\mathrm{k_{yv}}$ arise from phase matching and the Floquet conditions and are given by
\begin{eqnarray}
\mathrm{k_{xu}} &=&\mathrm{k_{0}\left[ n\times sin\theta cos\varphi -u\frac{\lambda }{\Lambda _{x}}\right] ,}  \label{ep5} \\
\mathrm{k_{yv}} &=&\mathrm{k_{0}\left[ n\times sin\theta sin\varphi -v\frac{\lambda }{\Lambda _{y}}\right] ,}  \label{ep6} \\
\mathrm{k_{z,uv}} &=&\mathrm{\left[ n^{2}k_{0}^{2}-k_{xu}^{2}-k_{yv}^{2}\right] ^{1/2}},\ \ \mathrm{n^{2}k_{0}^{2}\geq k_{xu}^{2}+k_{yv}^{2},}\label{ep7} \\
&=&\mathrm{-i\left[ k_{xu}^{2}+k_{yv}^{2}-n^{2}k_{0}^{2}\right] ^{1/2}},\ \ \mathrm{n^{2}k_{0}^{2}<k_{xu}^{2}+k_{yv}^{2}.}  \nonumber
\end{eqnarray}%
Here, $\mathrm{\Lambda }_{x}$ and $\mathrm{\Lambda }_{y}$ are the periodicity along the x and y directions, respectively; $\mathrm{\lambda }$ is the vacuum wavelength of the optical wave; $\mathrm{k_{0}=2\pi /\lambda }$ is the free space wave number; $\mathrm{n}$ is the refractive index of the incident/outgoing medium ($\mathrm{n}=1$ for the air in this work); $\mathrm{\theta }$ is the polar angle; and $\mathrm{\varphi }$ is the azimuth angle of the incident k-vector. We notice that the vector variables are in bold for convenience. On the basis of the spatial harmonics, Maxwell's equations can be expressed as an infinite set of first-order differential equations:
\begin{eqnarray}
\mathrm{\partial _{z}e_{x,uv}(z)} &=&\mathrm{-h_{y,uv}(z)+\frac{k_{xu}}{k_{0}^{2}}\sum_{m,n=-\infty }^{\infty }\overline{\varepsilon }_{u-m,v-n}\left( \ k_{xm}h_{y,mn}-k_{yn}h_{x,mn}\right) ,}  \label{mx1} \\
\mathrm{\partial _{z}e_{y,uv}(z)} &=&\mathrm{h_{x,uv}(z)+\frac{k_{yv}}{k_{0}^{2}}\sum_{m,n=-\infty }^{\infty }\overline{\varepsilon }_{u-m,v-n}\left( k_{xm}h_{y,mn}-k_{yn}h_{x,mn}\right) ,}  \label{mx2} \\
\mathrm{\partial _{z}h_{x,uv}(z)} &=&\mathrm{\frac{k_{xu}}{k_{0}^{2}}\left(k_{xu}e_{y,uv}-k_{yv}e_{x,uv}\right) -\sum_{m,n=-\infty }^{\infty}\varepsilon _{u-m,v-n}e_{y,mn}(z),}  \label{mx3} \\
\mathrm{\partial _{z}h_{y,uv}(z)} &=&\mathrm{\frac{k_{yv}}{k_{0}^{2}}\left(k_{xu}e_{y,uv}-k_{yv}e_{x,uv}\right) +\sum_{m,n=-\infty }^{\infty}\varepsilon _{u-m,v-n}e_{x,mn}(z).}  \label{mx4}
\end{eqnarray}

Together with specific incidence conditions, Eqs. (\ref{mx1})-(\ref{mx4}) can directly solve the vectorial fields $\mathrm{\mathbf{e}}$ and $\mathrm{\mathbf{h}}$ by means of eigenproblem algebra and can be easily extended to $\mathrm{\ell}$-level ($\mathrm{\ell}=2$ for the binary one) DOEs \cite{sa1,rcwa1}. For computationally efficient designs of DOEs, this work computes vectorial fields of an individual pixel under the pre-condition that the (singlepixel) field includes accurate characteristics of across-pixel interferences and sub-pixel fringes. In this way, KSAM can appeal to very finite k-series for still predictive computations, since the interference fields occur only among finite nearby pixels and the sub-pixel fringes are moderate under the condition $\mathrm{0\ll\Lambda _{ph}/\lambda \leq 1}$. By the reciprocity relations in Fourier algebra, we correlate the order of k-series with spatial parameters- i.e., the rank of neighboring interferences r, the sampling rate s, and the DOE depth h- for easily selecting the desired degrees of accuracy.

\begin{figure}[htbp]
\centering
\includegraphics[width=0.95\linewidth]{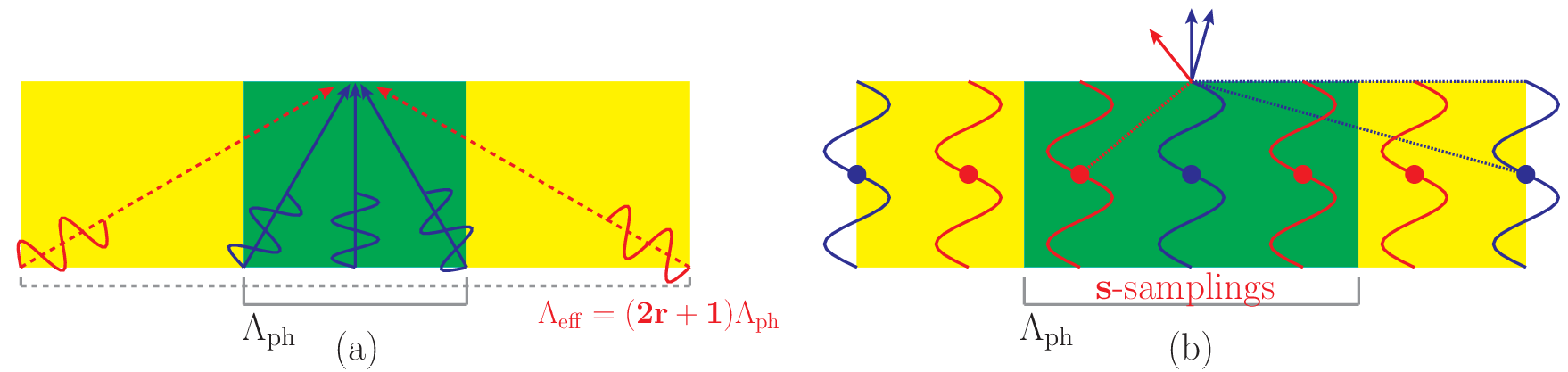}
\caption{(a) Schematic definition for the rank of neighboring interferences r=$(\Lambda_\mathrm{eff}/\Lambda_\mathrm{ph}-1)/2$ in the unit of pixel by r=1, and (b) schematic definition for the sampling rate s per pixel by s=3. }
\label{fig3x}
\end{figure}

Assuming negligible multiple reflections of electromagnetic waves, this work introduces an effective dimension $\mathrm{\Lambda_{eff}\equiv(2r+1)\Lambda_{ph}}$ to compute the fields for an assigned pixel, with the condition that the set-up can reproduce essential interferences' effects and simultaneously eliminate disturbances from non-physical periodicity in RCWA algebra. Moreover, since the spatial frequency $\mathrm{\varepsilon_{u-m,v-n}}$ of every stacking layer in DOEs can iteratively transfer wave vectors $\mathrm{\mathbf{e}_{mn}(\mathbf{h}_{mn})}$ into different diffraction components $\mathrm{\mathbf{e}_{uv}(\mathbf{h}_{uv})}$, as indicated in Eqs. (\ref{mx1})-(\ref{mx4}), simple criteria for spatial parameters can be given by $\mathrm{r\Lambda_{ph} \geq h k_{\parallel,R}/k_{z,R}}$ and $\mathrm{s \geq \ell -1}$. Here $\mathrm{k_{z,R}}$ is the selected highest-order k vector with a real-number z component, while $\mathrm{k_{\parallel,R}=(k_{x,R}^{2}+k_{y,R}^{2})^{1/2}}$ is the in-plane component of the k vector defined for $\mathrm{k_{z,R}}$. The total components of k-series are selected in the range $\mathrm{-(2r+1)s/2\leq u,v \leq(2r+1)s/2-1}$ with Eqs. (\ref{ep5})-(\ref{ep7}) at $\mathrm{\Lambda_{x(y)}=\Lambda_{eff}}$ in this work. The graphical interpretations of the optional parameters r and s are illustrated in Fig. \ref{fig3x}.

\begin{figure}[htbp]
\centering
\includegraphics[width=1.0\linewidth]{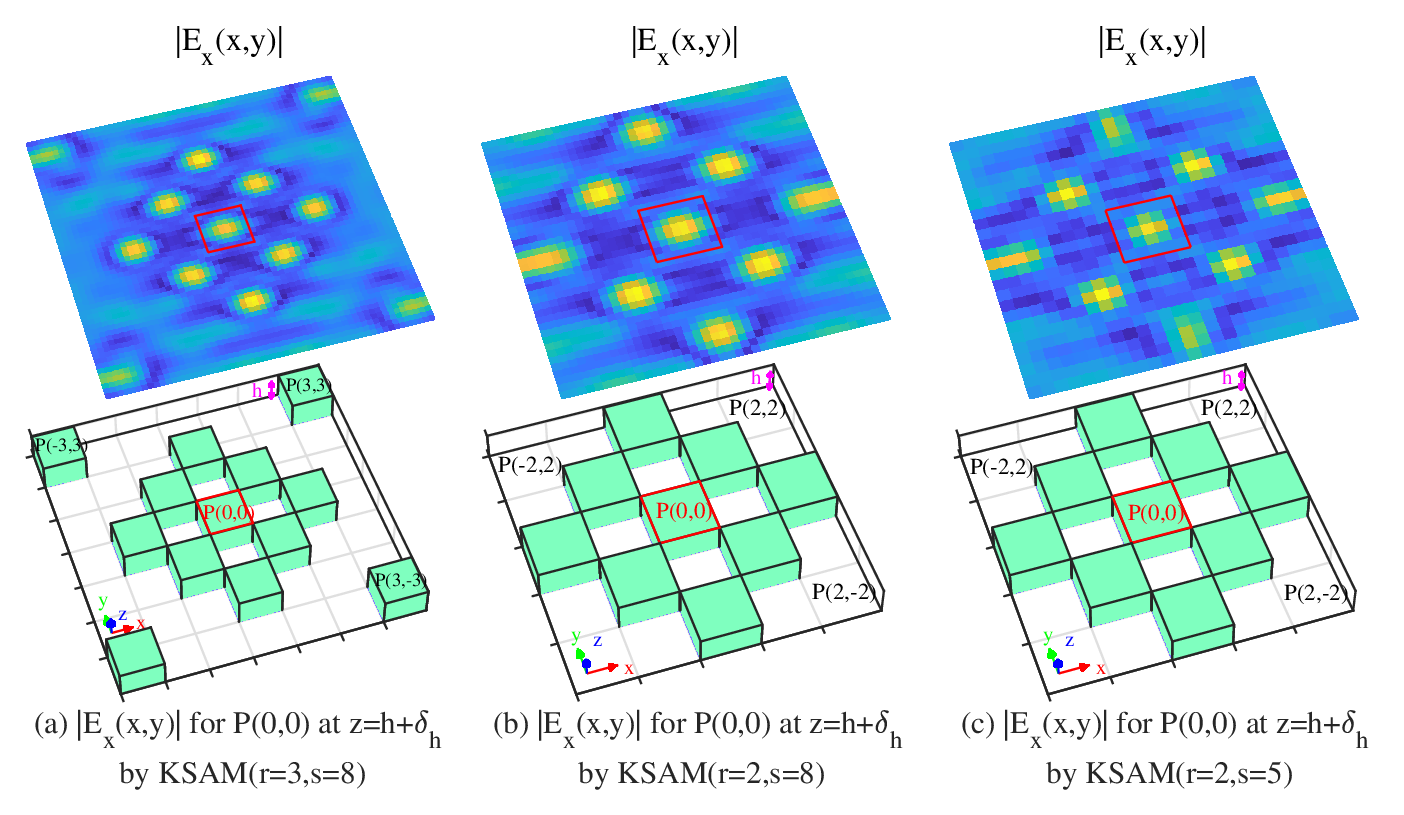}
\caption{Schematic examples for computing vectorial optical fields $\mathrm{|E_{x}(x,y)|}$ of the individual (red) pixel $\mathrm{P_{x}}$(0,0) by the k-series approximation method using different effective pixel $\mathrm{\Lambda_{eff}=(2r+1)\Lambda_{ph}}$ and sample rate s: (a) r=3 and s=8, (b) r=2 and s=8, and (c) r=2 and s=5.}
\label{fig4x}
\end{figure}

In Fig. \ref{fig3x}(a) the rank of neighboring interferences r defines the maximal pixel range of pixel-pixel interferences by $(\Lambda_\mathrm{eff}/\Lambda_\mathrm{ph}-1)/2$, where the size of an equivalent-isolated pixel $\mathrm{\Lambda_{eff}}$ should be infinite in theory for accuracy, but is set as a finite value by assuming insignificant interactions of fields among distant pixels. The computation hence requires only a finite morphology of nearby pixels within $\Lambda_\mathrm{eff}$ to determine the local field in $\Lambda_\mathrm{ph}$, but still can keep predictive calculations. In Fig. \ref{fig3x}(b) the sampling rate s defines the number of sampling points per pixel in one dimension. In this way, the k-series can represent an accurate sub-pixel profile of fields and enable occurrences of higher-order diffractions by reciprocity relations. With an adequate set-up of r and s values at given h, this work fulfills more efficient and accurate vectorial modeling of near fields in DOEs, in place of the typical wave front function in the scalar-based theory. Figure \ref{fig4x} illustrates schematic examples for computing the fields of the (red) pixel P(0,0) at different conditions (r,s) for KSAM, in which the lower part indicates the pixel morphology $\mathrm{\Upsilon}$ (with total dimension $\mathrm{\Lambda_{eff}}$ and $\mathrm{\Lambda_{ph}/\lambda=1}$), and the upper part shows the profiles of fields $\mathrm{|E_{x}(x,y)|}$ at $\mathrm{z=\lambda+\delta_{h}}$. Numerical results state that one can perform predictive optical designs with proper dimension r (comparing Figs. \ref{fig4x}(a) and \ref{fig4x}(b)) and sampling data s (comparing Figs. \ref{fig4x}(b) and \ref{fig4x}(c)), at an affordable computational cost ($\mathrm{92\%}$ reduction in computational time by a comparison of Figs. \ref{fig4x}(a) and \ref{fig4x}(c)). Here, the spatial domain size N is 1 for determining the (red) single-pixel fields by KSAM.

\begin{figure}[htbp]
\centering
\includegraphics[width=0.95\linewidth]{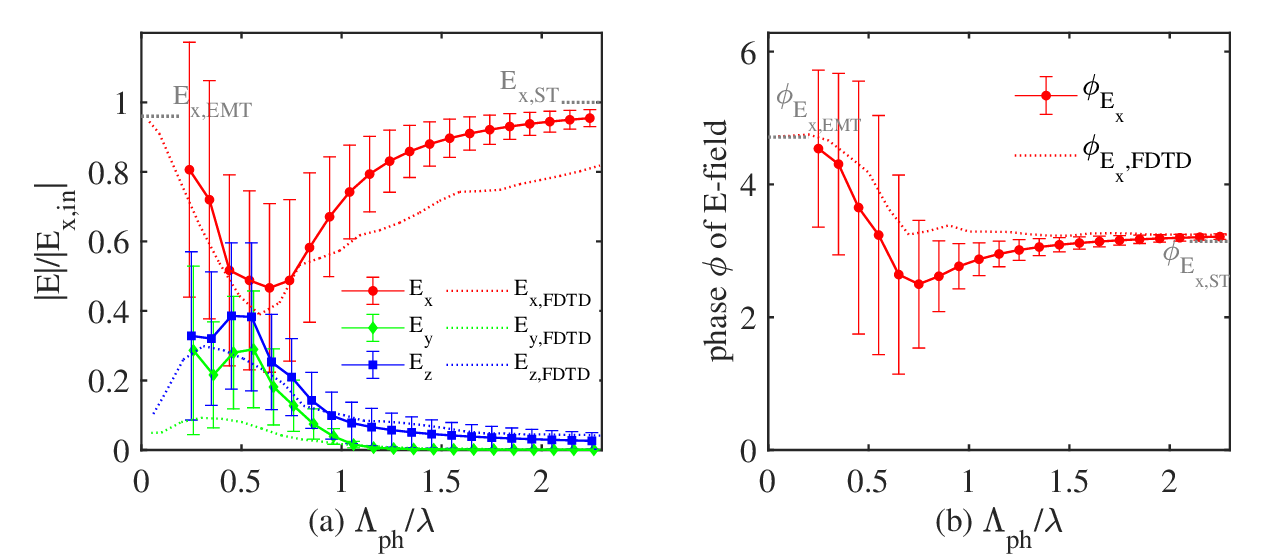}
\caption{(a) Normalized amplitude of different field components and (b) phase of the field's x-component at the central pixel as a function of pixel size by ensemble statistics. Solid curves indicate results by KSAM (r=s=2), and dotted curves show those by FDTD for comparison.}
\label{fig5x}
\end{figure}

\begin{figure}[htbp]
\centering
\includegraphics[width=0.95\linewidth]{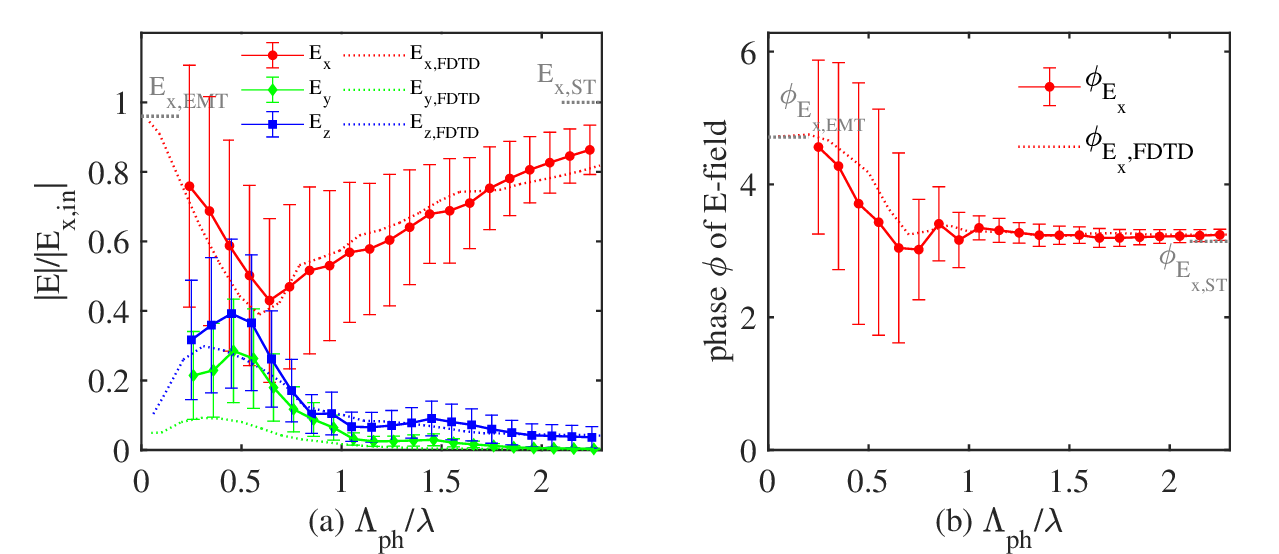}
\caption{(a) Normalized amplitude of different field components and (b) phase of the field's x-component at the central pixel as a function of pixel size by ensemble statistics. Solid curves indicate results by KSAM (r=s=4), and dotted curves show those by FDTD for comparison.}
\label{fig6x}
\end{figure}

To investigate the validity for general cases, we re-evaluate the analyses in Fig. \ref{fig2x} by KSAM and compare results with those by FDTD. Figure \ref{fig5x}(a) investigates the average amplitude of different field components at the specified (central) pixel using r=s=2. Figure \ref{fig5x}(b) plots the curve for the corresponding phase of x components of fields. Curves by FDTD are drawn by dotted lines for comparison. The asymptotic constant values based on the EMT and ST methods are denoted as horizontal dotted lines near the corresponding regimes, respectively. Numerical results in Fig. \ref{fig5x} display qualitatively similar variations of fields to that by FDTD when varying $\Lambda_\mathrm{ph}$. With higher-order (r=s=4) approximations, Fig. \ref{fig6x} shows quantitatively similar curves to that by FDTD and clearly predict more accurate vectorial effects (e.g., depolarization and dephasing of fields) for systems studied.

\subsection{Vector-based simulated annealing algorithm for designing diffractive optical elements}

To next interpret the process for optical designs, this work exemplifies a Fourier-transformed DOE, where the coordinates of the image plane are denoted ($\mathrm{x_{I}}$,$\mathrm{y_{I}}$) and those of the exiting plane on DOE are denoted ($\mathrm{x_{D,s}}$,$\mathrm{y_{D,s}}$). DOE is assumed to be illuminated by a plane wave. The reconstructed image is derived in the focal (image) plane of the Fourier lens.

During the design process, trial morphologies $\mathrm{\Upsilon'(x_{D},y_{D})}$ are specifically first generated with a simulated annealing algorithm, and the pixel-field profiles $\mathrm{\mathbf{P}'(\mathrm{x_{D,s}},\mathrm{y_{D,s}})}$ are then computed by means of KSAM. Here, the subscript s represents the index of sampling points per pixel. The DOE quality hence is scored by a cost function associated with the diffraction image of $\mathrm{\mathbf{P}'}$ and the target image, in which the diffraction fields $\mathrm{R_{x(y)}}$ in the image plane are given by
\begin{eqnarray}
&&\mathrm{R_{x(y)}(x_{I},y_{I})} \subset\mathrm{F\left[ P_{x(y)}(x_{D,s},y_{D,s})\right]^{-1} } \label{vss1} \\
&=&\mathrm{\sum_{x_{D},y_{D}=-N/2}^{N/2-1}\sum_{p_{i},p_{j}=1}^{s}P_{x(y)}(x_{D,p_{i}},y_{D,p_{j}})e^{2\pi i\left( \frac{x_{D,p_{i}}}{sN}x_{I}+\frac{y_{D,p_{j}}}{sN}y_{I}\right) }.}  \nonumber
\end{eqnarray}
Here, $\mathrm{F^{-1}}$ is the inverse Fourier transform, and the $\mathrm{R_{z}}$ component is obtained with the relation $\mathrm{\mathbf{k}(x_{I},y_{I})\cdot\mathbf{R}(x_{I},y_{I})=0}$. Moreover, DOE is considered as an optical element of $\mathrm{N^2}$ square pixels, and the pixel-field profile $\mathrm{\mathbf{P}}$ is an $\mathrm{(N\times s)^2}$ array function. In this work $\mathrm{x_{I}}$ and $\mathrm{y_{I}}$ are simply assigned as the lowest $\mathrm{N^{2}}$-order wave vectors of the total $\mathrm{(N\times s)^2}$ ones, responding to an $\mathrm{N^{2}}$-pixel target image $\mathrm{I_{0}}$. The intensity I of the reconstructed image is given by:
\begin{equation}
\mathrm{I(x_{I},y_{I})}=\mathrm{\left\vert \mathbf{R}(x_{I},y_{I})\right\vert^{2}.} \label{vss2}
\end{equation}
The cost or energy function to score performance of reconstruction is defined as the mean-square error between the reconstructed image and the target images $\mathrm{I_{0}}$, presented by:
\begin{equation}
\mathrm{m}=\mathrm{\int\int\left\vert I_{0}(x_{I},y_{I})-\alpha I(x_{I},y_{I})\right\vert ^{2}dx_{I}dy_{I}.} \label{vss3}
\end{equation}
Here, $\alpha$ is a scale factor given by:
\begin{equation}
\mathrm{\alpha }=\mathrm{\frac{\int \int I_{0}(x_{I},y_{I})dx_{I}dy_{I}}{\int \int I(x_{I},y_{I})dx_{I}dy_{I}}.} \label{vss4}
\end{equation}
We note that the cost function is obviously zero when the reconstructed image equals the desired image.

\begin{figure}[htbp]
\centering
\includegraphics[width=1.0\linewidth]{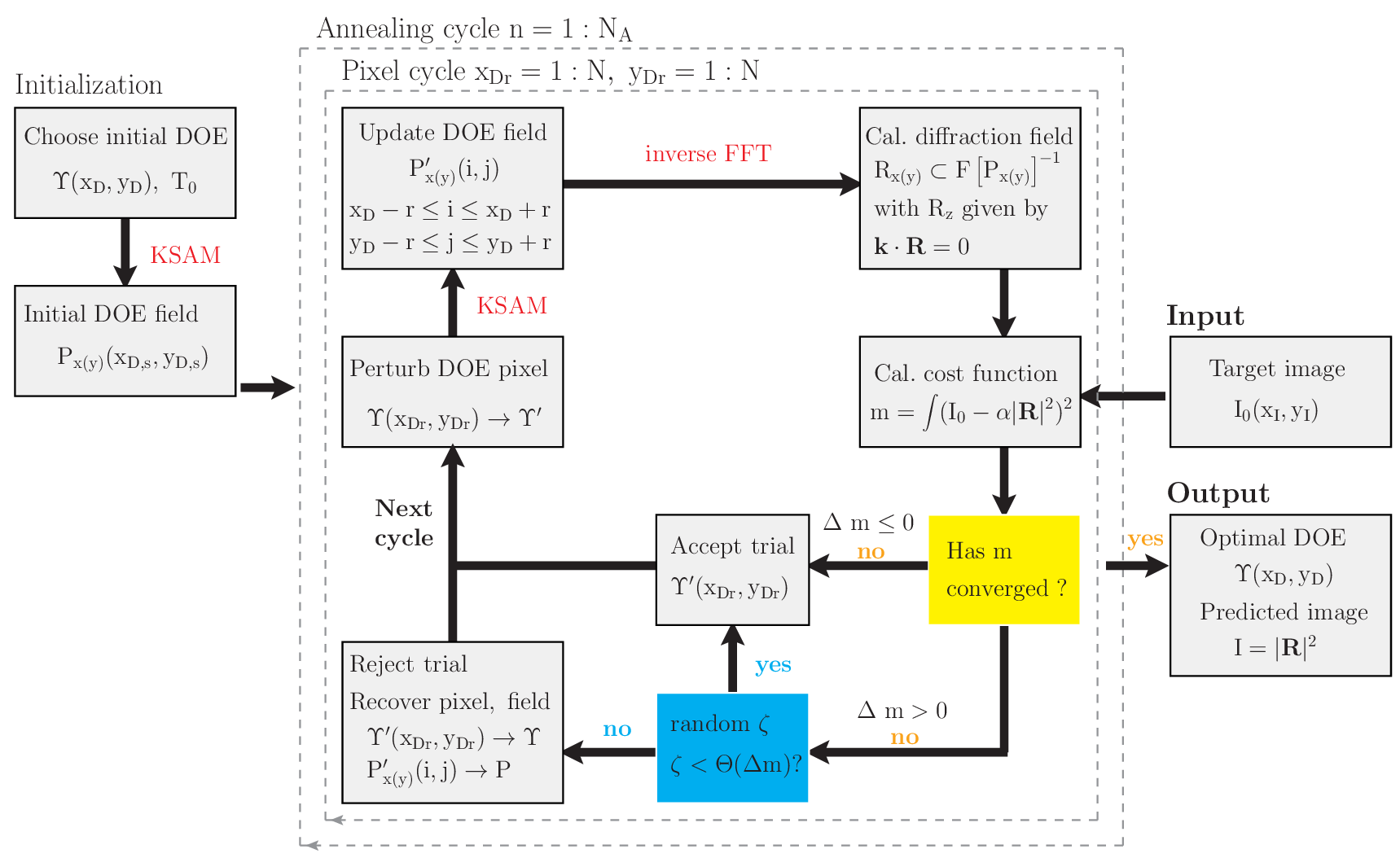}
\caption{Schematic flow chart illustrating the vector-based simulated annealing algorithm for diffractive optical elements with subwavelength feature sizes and for intensity objects }
\label{fig7x}
\end{figure}

The vector-based simulated annealing algorithm can be summarized by the following steps, and the flow chart is plotted in Fig. \ref{fig7x}.
\begin{enumerate}
\item The pixel morphology $\mathrm{\Upsilon(x_{D},y_{D})}$ of DOE is initialized; e.g., according to the discretization of phase distribution of the Fourier transform of a desired image $\mathrm{I_{0}}$. The corresponding pixel-field profile $\mathrm{\mathbf{P}(x_{D,s},y_{D,s})}$ is then initially built by KSAM. The temperature parameter $\mathrm{T_{0}}$ of the simulated annealing process is initially set at a relatively high value, enabling a large range of perturbation probability.
\item Each pixel $\mathrm{\Upsilon(x_{Dr},y_{Dr})}$ of DOE is successively selected for fluctuation during the pixel cycle ($\mathrm{x_{Dr}=1:N}$ and $\mathrm{y_{Dr}=1:N}$). The morphology (or depth) of this selected pixel is reset by a random (trial) candidate of depth levels. The fields of pixel $\mathrm{\mathbf{P}'(i,j)}$ in the range ($\mathrm{x_{D}-r\leq i\leq x_{D}+r}$,$\mathrm{y_{D}-r\leq j\leq y_{D}+r}$) are then updated by KSAM, where these pixels belonging to device boundaries are assumed to follow a periodic arrangement for simplification. We note that updating multiple pixel fields $\mathrm{\mathbf{P}'(i,j)}$ for a single trial pixel $\mathrm{\Upsilon(x_{Dr},y_{Dr})}$ here can bring forth a factor of $\mathrm{(2r+1)^{2}s^{2}}$ for the computation time.
\item Calculate the diffraction fields by Eq. (\ref{vss1}). The difference in the cost function $\mathrm{\Delta m}$=$\mathrm{m_{new}}$-$\mathrm{m_{now}}$ is then calculated by Eqs. (\ref{vss2})-(\ref{vss4}). If $\mathrm{\Delta m}$ is negative, then the new pixel morphology is accepted; otherwise, acceptance or rejection is determined by the Boltzmann probability:
\begin{equation}
\mathrm{\Theta (\Delta m,T)=e^{-\frac{\Delta m}{T}},} \label{fc1}
\end{equation}
    where a random number $\mathrm{0\leq \zeta\leq 1}$ is first generated, and the new (trial) pixel is accepted if $\mathrm{\zeta<\Theta (\Delta m,T)}$ occurs and is rejected otherwise.
\item Repeat steps 2 and 3 for the pixel cycle ($\mathrm{x_{Dr}}$ and $\mathrm{y_{Dr}}$) and the annealing cycle ($\mathrm{n=1:N_{A}}$); except when the criterion of convergence is achieved, where T in Eq. (\ref{fc1}) is made lower with the number of annealing cycles by:
\begin{equation}
\mathrm{T=\frac{T_{0}}{1+n}}, \ \ \mathrm{1\leq n \leq N_{A}.} \label{fc2}
\end{equation}
\end{enumerate}
An alternate Fourier algorithm for free-space propagation of vectorial optical fields, compared to Eq. (\ref{vss1}), can be seen in \cite{fftv1}. The Matlab code concerning the first-order algorithm (r=1 and s=1) for designs of square DOEs is available in MATLAB Central File Exchange \cite{matlab1}.

\subsection{Two-stage optimization process for high-quality DOEs}

\begin{figure}[htbp]
\centering
\includegraphics[width=0.85\linewidth]{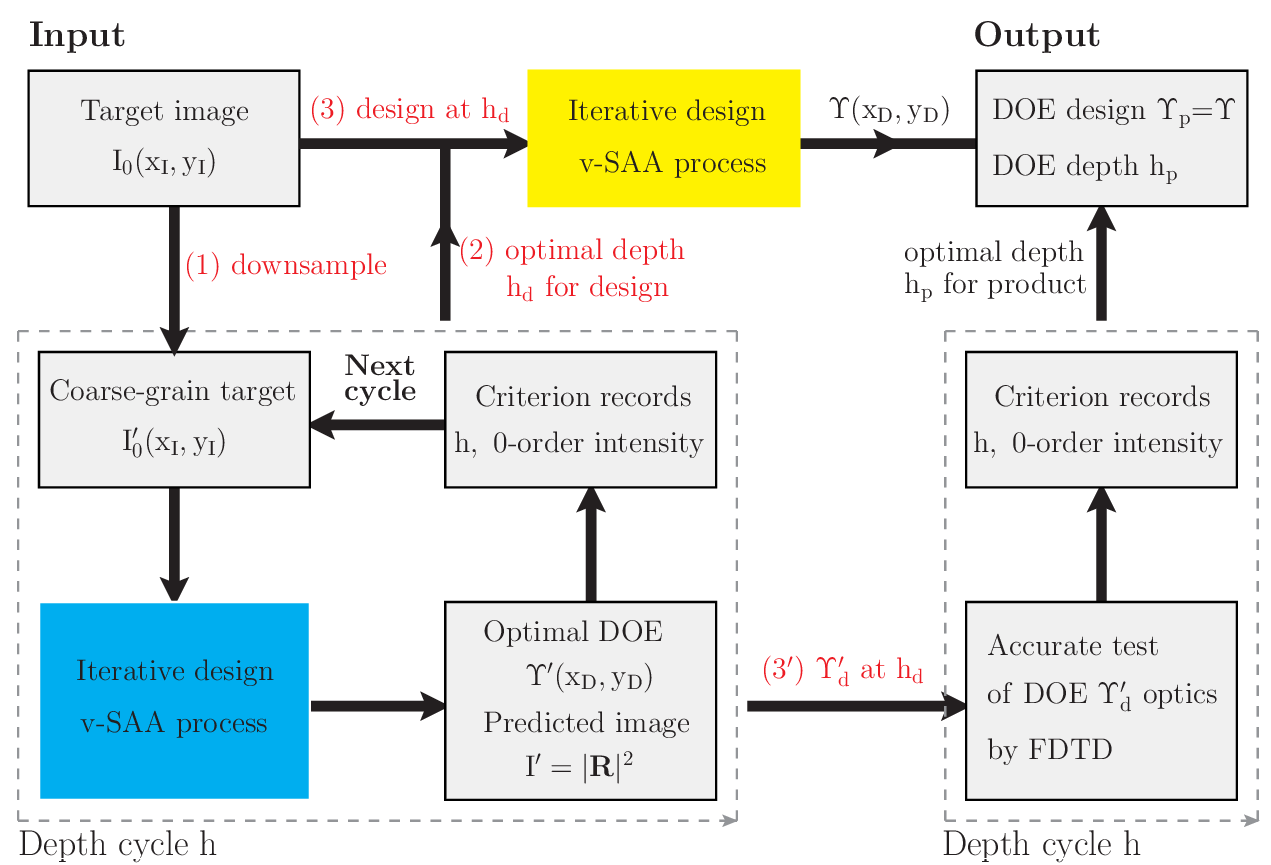}
\caption{Block diagram of optimization processes for the depth dimension h of DOEs, on the basis of v-SAA modules (colored blocks). }
\label{fig8x}
\end{figure}

To achieve high-quality DOEs, we introduce complementary optimization processes for deviations by hypothetical approximations. Specifically, for systems with microstructures at the subwavelength scale, field errors are primarily attributed to two issues: vectorial effects and approximation by the truncation of k-series. Thus, a two-stage optimization process is proposed in this work to deal with these two issues. To enable fast estimations, we employ a scaled-down sample of the desired product as a temporary prototype during the two-stage process. A schematic flow diagram is plotted in Fig. \ref{fig8x}. Referring to Fig. \ref{fig2x}, the vectorial optical fields in the studied regime $\mathrm{0\ll\Lambda _{ph}/\lambda \leq 1}$ noticeably depart from the prediction by the scalar-based theory. The optimized DOE depth, binary DOE for instance, could hence differ from the ideal value as well. For this reason, it is necessary to identify the optimal depth h by the vector-based theory in the first stage. By applying separate DOE designs $\mathrm{\Upsilon'}$ at different depth value, the highest-quality one $\mathrm{\Upsilon'_{d}}$ and the optimized design depth $\mathrm{h_{d}}$ are determined, as processes (2) and (3$\mathrm{'}$) in Fig. \ref{fig8x}. Here, the popular specification by the 0-order diffraction is adopted as the quality criterion. In the second stage, since the transmission fields by low-order KSAM deviate from that by the rigorous FDTD method (see Figs. \ref{fig5x} and \ref{fig6x}), the design depth $\mathrm{h_{d}}$ in theory can be inconsistent with the optimized value $\mathrm{h_{p}}$ for products. To determine the optimal $\mathrm{h_{p}}$ value, the design $\mathrm{\Upsilon'_{d}}$ is scored as a function of depth h by FDTD. It is concluded that the final DOE product $\mathrm{\Upsilon_{p}}$ is designed with v-SAA at $\mathrm{h_{d}}$ and is manufactured with the set-up of depth $\mathrm{h_{p}}$ for optimal performances. One can take it for granted that the value $\mathrm{h_{d}}$ will asymptotically converge to the value $\mathrm{h_{p}}$ in the high-order limit of the k-series approximation.

\section{Design results and analysis}

\begin{figure}[htbp]
\centering
\includegraphics[width=0.95\linewidth]{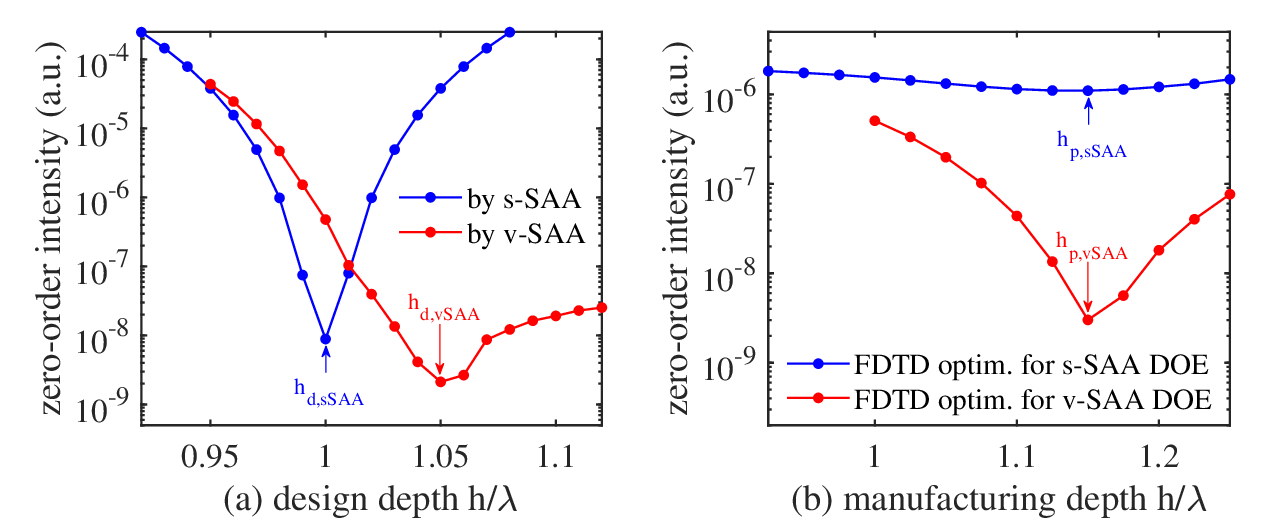}
\caption{(a) The intensity of zero-order diffraction corresponding to different DOEs that are separately designed with varying $\mathrm{h/\lambda}$, using s-SAA and v-SAA models, respectively, and (b) Revaluations of zero-order intensity as a function of $\mathrm{h/\lambda}$ for candidate DOE in (a) by means of FDTD, which determines the final optimal product depth for s-SAA and v-SAA models, respectively.}
\label{fig9x}
\end{figure}

\begin{figure}[htbp]
\centering
\includegraphics[width=0.95\linewidth]{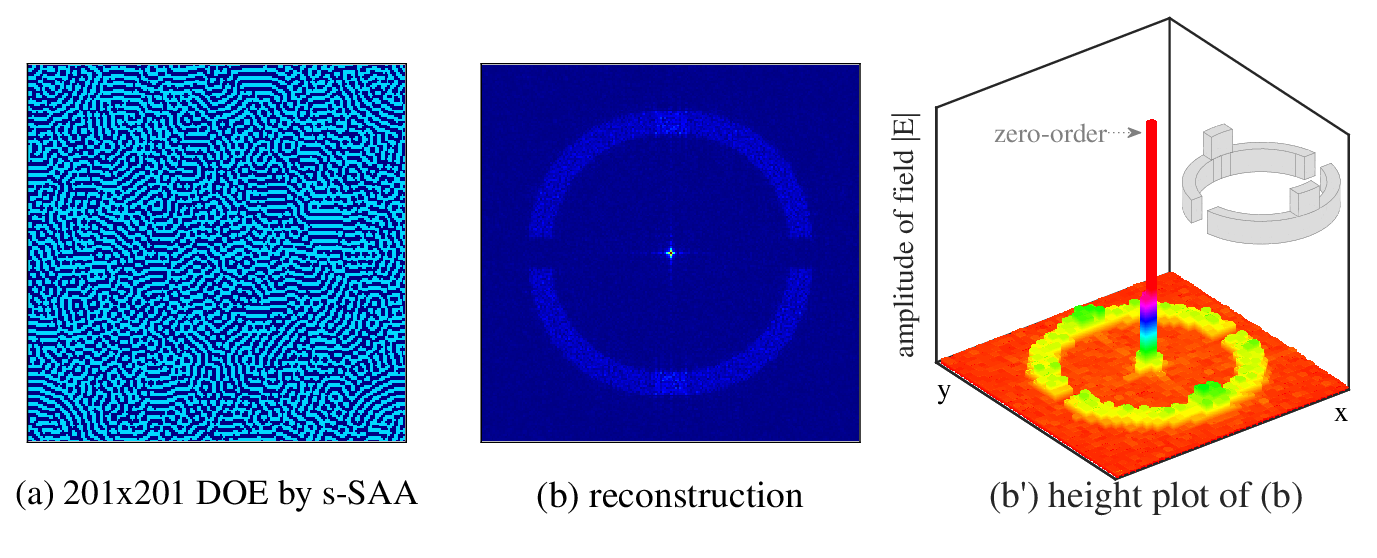}
\caption{(a) The binary morphology of DOE designed by s-SAA, (b) the reconstructed image, and (b') its height plot in lower resolution. }
\label{fig10x}
\end{figure}

\begin{figure}[htbp]
\centering
\includegraphics[width=0.95\linewidth]{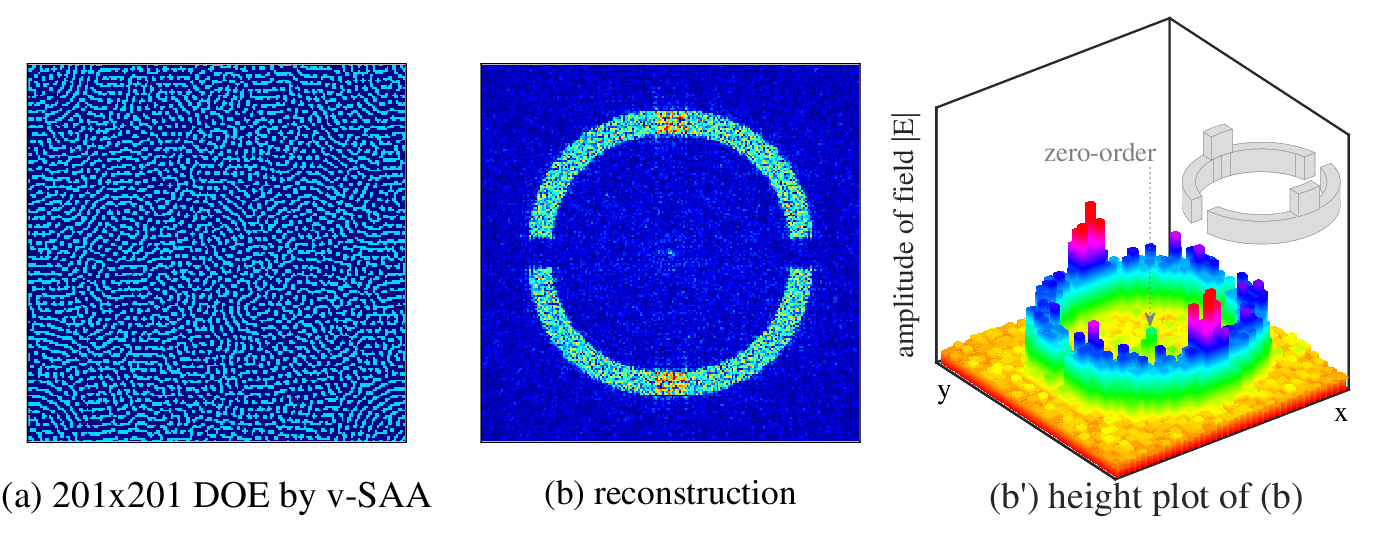}
\caption{(a) The binary morphology of DOE designed by v-SAA (r=1, s=1), (b) the reconstructed image, and (b') its height plot in lower resolution.}
\label{fig11x}
\end{figure}

To evaluate the feasibility of the proposed algorithm, a beam shaper is adopted as an exemplary target. The simulation conditions are set by the plane-wave incidence, the wavelength $\mathrm{\lambda=1.0} \ \mathrm{\mu m}$, the incident angle $\mathrm{\theta=0^{\circ}}$, TM polarization, the refractive index of air $\mathrm{n_{1}=1.0}$, the refractive index of DOE material $\mathrm{n_{2}=1.5}$, the physical pixel size $\mathrm{\Lambda_{ph}=0.6\mu m}$, the rank of neighboring interferences r=1, and the sampling rate s=1 position per pixel, for designing a $\mathrm{201\times201}$ pixel array of binary DOEs. The default depth is set as $\mathrm{h_{def}=1.0\lambda}$ by s-SAA, and its optimal value is re-evaluated by the two-stage optimization process as in Fig. \ref{fig9x}. A 3D schematic of DOEs is shown in Fig. \ref{fig1x}(c). The full structure of DOEs is represented by an image matrix, as in Figs. \ref{fig10x}(a) and \ref{fig11x}(a), where the cyan area indicates the material medium ($\mathrm{n_{2}}$), and the blue area is the air-filling region ($\mathrm{n_{1}}$). The beam shaper target is indicated as the inset mesh in Figs. \ref{fig10x}(c) and \ref{fig11x}(c).

Before generating the full $\mathrm{201\times201}$ design, a temporary $\mathrm{101\times101}$ DOE is employed for speeding up the two-stage optimization process, so as to quickly determine the optimal design depth $\mathrm{h_{d}}$ and product depth $\mathrm{h_{p}}$. At the first stage, with varying depth parameter h, a different DOE morphology is separately generated, and its intensity of zero-order diffraction is recorded for a quality score. Numerical results in Fig. \ref{fig9x}(a) show that the optimal design depths are $\mathrm{h_{d,sSAA}=\lambda}$ and $\mathrm{h_{d,vSAA}=1.05\lambda}$, corresponding to the binary morphologies $\mathrm{\Upsilon'_{d,sSAA}}$ and $\mathrm{\Upsilon'_{d,vSAA}}$ by s-SAA and v-SAA models, respectively. At the second stage, with given designs ($\mathrm{\Upsilon'_{d,sSAA}}$ and $\mathrm{\Upsilon'_{d,vSAA}}$), the zero-order intensities at different set-up of depths are directly estimated by FDTD, in order to determine the optimal depth $\mathrm{h_{p}}$ for products. Figure \ref{fig9x}(b) shows that the optimal product depths are $\mathrm{h_{p,sSAA}=h_{p,vSAA}=1.15\lambda}$ for this case. Here, the large zero-order intensity signifies the low-quality DOEs that are designed by the s-SAA model at a subwavelength feature scale.

With the known optimal depth parameter $\mathrm{h_{d}}$, the full $\mathrm{201\times201}$ morphology of DOE can then be generated by SAA. Figure \ref{fig10x}(a) illustrates the binary morphology of DOE designed by the s-SAA model with $\mathrm{h_{d}=1.0\lambda}$. The 4-thread computation over 100 annealing cycles runs in 160.9 seconds, which converges the variation of the cost function to below $\mathrm{0.01\%}$ per iteration. The $\mathrm{201\times201}$ morphology of DOE designed by the v-SAA(r=1,s=1) model with $\mathrm{h_{d}=1.05\lambda}$ is depicted in Fig. \ref{fig11x}(a). Similarly, the 4-thread optimization computation over 100 cycles runs in 1336.9 seconds, plus an extra time of 1.4 seconds for building the lookup table having $\mathrm{\ell^{(2r+1)^{2}}s^{2}}$-size. To validate the designs, with given $\mathrm{h_{p}=1.15\lambda}$, the reconstructed images of DOEs are directly simulated by means of well-known softwares \cite{fdtd1,fdtd2} that use the rigorous electromagnetic theory FDTD. Figure \ref{fig10x}(b) shows the reconstructed image that corresponds to the design in Fig. \ref{fig10x}(a), and its height plot in lower resolution is shown in Fig. \ref{fig10x}(b'). The target pattern is put as an inset mesh in Fig. \ref{fig10x}(b') for comparison. Here, rather than the typical intensity image $\mathrm{I=|E|^2}$, the reconstructed image is presented by the amplitude value of fields $\mathrm{|E|}$ for clarity. The FDTD program by 4-threads estimates the optics of DOE in 23389.1 seconds, in comparison with 1.6 seconds by s-SAA and 13.4 seconds by v-SAA. Numerical results indicate that the DOE designed by the s-SAA model exhibits low diffraction efficiencies, owing to the overwhelming zero-order intensity. Numerical results for the design by v-SAA are reported in Fig. \ref{fig11x}. The corresponding reconstructed images are illustrated in Fig. \ref{fig11x}(b) and \ref{fig11x}(b'). Accordingly, we find that diffraction efficiencies as well as the design quality can be significantly enhanced by the vector-based method. Beyond the mentioned cases, designs with a finite device extent of, multi-level ($\mathrm{\ell\geq3}$), and/or multiple functional diffractive optical elements could also be generated by the presented model.

\section{Conclusion}

This research has described in detail a vector-based optimization tool for designing DOEs with subwavelength feature sizes. By introducing an adequate selection scheme of k-series, the algorithm can perform computationally efficient yet with predictive computations. Together with an optimization of the geometrical degrees of freedom (i.e., DOE depth h) as compensation for errors from the truncation of k-series, this work presents DOEs with diffraction efficiencies close to the theoretical limit. Furthermore, after implementing with lookup table techniques to pre-calculate pixel fields under ergodic couplings with nearby ones, the vector-based algorithm can achieve large-extent (mm- to cm-scale) designs in time frames comparable to those of scalar-based models.

\begin{backmatter}

\bmsection{Acknowledgments}
This work was supported by Institute of Nuclear Energy Research (INER) under project AIE010403.

\bmsection{Disclosures}
The authors declare no conflicts of interest.

\bmsection{Data availability} The Matlab code concerning the first-order algorithm (r=1 and s=1) in this paper is available in Ref. \cite{matlab1}.

\end{backmatter}

\end{document}